\documentclass[prb,preprint,showpacs,showkeys]{revtex4}

\usepackage{hyperref}
\usepackage[english]{babel}
\usepackage[T1]{fontenc}
\usepackage[latin1]{inputenc}
\usepackage[dvips]{graphicx}
\usepackage{amsmath}
\usepackage{mathenv}
\usepackage{array}
\usepackage{bm}
\usepackage{fancyhdr}
\usepackage{makeidx}
\usepackage{subfigure}
\usepackage{babel}
\usepackage[v2]{xy}
\usepackage{epsfig}
\usepackage{amsfonts}
\usepackage{amssymb}
\usepackage{color}
\hypersetup{backref=true,pagebackref=true,hyperindex=true,colorlinks=true,breaklinks=true,urlcolor=
blue,linkcolor=blue,bookmarks=true,bookmarksopen=true}

\begin{document}

\title{Analysis of anisotropy crossover due to oxygen in Pt/Co/MOx trilayer}
\date{\today}
\author{A. Manchon}
\affiliation{SPINTEC, URA 2512 CEA/CNRS, CEA/Grenoble, 38054 Grenoble Cedex 9, France}
\author{S. Pizzini}
\author{J. Vogel}
\author{V. Uhl\' i\v r}
\affiliation{Institut Néel, CNRS/UJF, B.P. 166, 38042 Grenoble Cedex 9, France}
\author{C. Ducruet}
\author{L. Lombard}
\author{S. Auffret}
\author{B. Rodmacq}
\author{B. Dieny}
\affiliation{SPINTEC, URA 2512 CEA/CNRS, CEA/Grenoble, 38054 Grenoble Cedex 9, France}
\author{M. Hochstrasser}
\affiliation{Laboratory for Solid State Physics, ETH Zürich, 8093 Zürich, Switzerland}
\author{G. Panaccione}
\affiliation{Laboratory TASC, INFM-CNR, Area Science Park, S.S.14, Km 163.5, I-34012, Trieste, Italy
}\begin{abstract}
Extraordinary Hall effect and X-ray spectroscopy measurements have been performed on a series of Pt/Co/MOx trilayers (M=Al, Mg, Ta...) in order to investigate the role of oxidation in the onset of perpendicular magnetic anisotropy at the Co/MOx interface. It is observed that varying the oxidation time modifies the magnetic properties of the Co layer, inducing a magnetic anisotropy crossover from in-plane to out-of-plane. We focused on the influence of plasma oxidation on Pt/Co/AlOx perpendicular magnetic anisotropy. The interfacial electronic structure is analyzed via X-ray photoelectron spectroscopy measurements. It is shown that the maximum of out-of-plane magnetic anisotropy corresponds to the appearance of a significant density of Co-O bondings at the Co/AlOx interface.
\end{abstract}
\maketitle
\maketitle
\clearpage

\section{Introduction}
The recent demonstration of current-induced magnetization excitations in magnetic tunnel junctions \cite{huai,fuchs} and its potential application in magnetic random access memories\cite{ieee} (MRAM) has shown the necessity to fabricate high quality low resistance magnetic tunnel junctions (MTJ). Since the first publication of room temperature tunnelling magnetoresistance \cite{moodera} (TMR), the development of deposition and oxidation techniques allowed research groups to reach very high TMR ratios up to 100\% in AlOx-based MTJs \cite{alox} and 500\% in MgO-based MTJs\cite{ieee}. However, controlling the oxidation process in micronic devices imposes to achieve heavy lithographic processes. In-situ TMR characterisation would be of great interest to get rid of this microfabrication step.\par
To do so, Worledge et al.\cite{worl} proposed an original way to estimate the junctions' TMR from current-in-plane measurements, without using any lithographic patterning of these junctions. Unfortunately, this technique becomes harder to use when the MTJ resistance-area product becomes smaller. Rodmacq et al.\cite{monso, rodmacq} recently proposed another technique to control the oxidation of an insulating barrier, correlating the oxygen-induced perpendicular magnetic anisotropy (PMA) of a trilayer composed of Pt/Co/MOx (M is a metal - Al, Mg, Ta...) with the maximum of TMR of the corresponding Co/MOx/Co MTJ as a function of the barrier oxidation.\par

However, the role of oxygen in this "anisotropy crossover" has not yet been fully understood. In this article, we propose an investigation of the role of oxygen in the enhancement of the interfacial magnetic anisotropy of a thin cobalt layer sandwiched between a platinum layer and a non-magnetic oxide. Magnetic measurements are correlated to results of X-ray Absorption Spectroscopy (XAS) and X-ray Photoelectron Spectroscopy (XPS).\par
The paper is organized as follows : after a brief introduction on the role of interfacial states in PMA and TMR (section \ref{s:2}), section \ref{s:3} presents the conditions of sample preparation. The magnetic properties are described in section \ref{s:4} and an analysis of X-ray measurements is presented in section \ref{s:5}. We conclude the article in section \ref{s:7}.

\section{Interfacial states and perpendicular magnetic anisotropy\label{s:2}}
\subsection{Electronic band structure effects at the Co/AlOx interface}

Systems consisting of F/MOx interfaces (F is a ferromagnet and M is a diamagnetic metal, MOx being a non-magnetic oxide) have attracted much interest in the scientific community from both magnetic and electronic points of view\cite{regan,reso}. Tsymbal et al. \cite{reso,etatsreso} have theoretically demonstrated that in magnetic tunnel junctions, interfacial densities of states, which determine the amplitude of TMR, can be strongly influenced by localized interfacial atoms. Oleinik et al. \cite{oleinik} have theoretically studied the role of interfacial oxygen in Co/AlOx interfaces in different configurations (see also Ref.~\onlinecite{stoef}). When AlOx is under-oxidized, the interface termination is composed of Co-Al-O bondings: a charge transfer occurs between Co and O through Al and reduces the magnetic moment of Co, reducing the TMR amplitude. When the AlOx barrier is optimally oxidized\cite{belaprb}, the interface termination is Co-O-Al: hybridization between Co 3$d$ and O 2$p$ bands strongly modifies the band structure of both Co and O, so that O behaves like a magnetic insulator, enhancing the TMR amplitude. Belaschenko et al.\cite{belaprb} predicted that "adsorbed" oxygen near the Co/AlOx interface leads to positive spin polarization of the tunnelling current. These theoretical results underline the dominant role of oxygen atoms in the band structure of Co at the Co/AlOx interface: the precise chemical composition of the interface controls the electrical properties of the MTJ.\par

Telling et al. \cite{telling2004,telling2006} have experimentally demonstrated the close relation between AlOx oxidation, Co magnetic moment and TMR, showing that a maximum of TMR is reached when the Co magnetic moment is maximum, upon optimal oxidation. The authors then proved that the study of the magnetic properties of Co can give information on the electronic band structure at the interfaces, and thus monitoring the Co magnetic properties can be used to control the barrier oxidation.\par

\subsection{Perpendicular Magnetic Anisotropy}

The discovery of perpendicular magnetization anisotropy (PMA) in Pd/Co, Pt/Co and Au/Co multilayers \cite{carcia85} has opened an exciting field of research questioning the fundamental origin of such PMA and the role of interfacial orbital hybridization\cite{daalderop,nakajima}. Ferromagnetic materials usually used in these studies are Fe, Co and Ni whose atomic structure is 3$d^n4s^2$ (n=6, 7, 8 resp.). The localized $3d$ electrons being responsible for the spontaneous magnetization of these ferromagnets, the source of PMA is related to the anisotropy of these orbitals. In a bulk ferromagnet, the 3$d$ orbitals are degenerate in a first approximation (the chemical environment is roughly spherical). The presence of an interface breaks this quasi-spherical symmetry so that the energy of the 3$d$ orbitals pointing towards the interface ($d_{xz}$, $d_{yz}$ and $d_{z^2}$) is different from the energy of the 3$d$ orbitals with planar symmetry ($d_{xy}$ and $d_{x^2-y^2}$).\par
This gives rise to a crystalline field $\Delta_c=E(yz, zx, z^2)-E(xy, x^2-y^2)$. Bruno\cite{bruno} showed that the magnetic anisotropy energy $\Delta E$ is strongly influenced by this parameter and can be written as:
\begin{eqnarray}
\Delta E=\frac{G}{H}\frac{\xi}{4\mu_B}(m_{orb}^\bot-m_{orb}^{||})
\end{eqnarray}
where $m_{orb}^{\bot(||)}$ is the out-of-plane (in-plane) orbital moment, $\xi$ is the spin-orbit coupling parameter and $G/H$ are band-structure dependent coefficients. The magnetic anistropy arises from the coupling between the anisotropy of the 3$d$ orbital moment $\overrightarrow{l}$ and the magnetic moment $\overrightarrow{s}$ of the ferromagnetic atoms through the spin-orbit coupling $\xi\overrightarrow{s}.\overrightarrow{l}$. However, spin-orbit coupling is rather difficult to estimate, its energy being of the order of a few meV.\par

Daalderop et al. \cite{daalderop} have studied the magnetic anisotropy energy (MAE) in a crystalline Co monolayer. The authors demonstrated that in this case, the MAE changes its sign as a function of the filling of the Co valence band. This was confirmed by Kyuno et al. \cite{kyuno} on Co/X (X=Pt, Pd, Cu, Ag, Au) interfaces. This band filling can be tuned by hybridization between the Co 3$d$ orbitals and the orbitals of layer X. For example, in the case of Co/Pt interface, Nakajima et al. \cite{nakajima} demonstrated that the 3$d$ bands of cobalt are hybridized with the 5$d$ bands of platinum.\par

Furthermore, Weller et al. \cite{weller} showed that in the case of strong spin-orbit coupling (like in the case of Co/Au interfaces), this spin-orbit coupling is sufficiently large to align the large spin moment parallel to the small orbital moment. Then the MAE can be modified by the band splitting of the 3$d$ orbitals due to the coupling with the local environment, and by the amplitude of the spin-orbit coupling itself.\par

Consequently, at the Co/AlOx interfaces, one can expect that the predicted charge transfer\cite{reso,oleinik,belaprb} between Co and O in optimally oxidized Co/AlOx interfaces increases the asymmetry of the Co 3$d$ bands, reducing the energy of the 3$d$ orbitals responsible for the out-of-plane anisotropy and creating a splitting between in-plane and out-of-plane $d$ orbitals (parameter $\Delta$ in Bruno's theory \cite{bruno}). Thus, in spite of the weak spin-orbit coupling of O atoms, this strong band splitting could lead to measurable PMA.

\section{Sample Preparation\label{s:3}}

In order to study the influence of oxidation on the Co magnetization, Pt(3 nm)/Co(0.6 nm)/M+Oxidation (M=Al, Ta, Mg, Ru) were deposited on a thermally oxidized silicon wafer by conventional dc magnetron sputtering with a base pressure of $10^{-8}$ mbar. Two oxidation techniques were used: some of the samples were oxidized naturally under a pressure of 160 mbar during 30 minutes. These samples were generally annealed at 150°C during 30 minutes. Other samples were oxidized using an oxygen rf plasma with a partial pressure of $3\times10^{-3}$ mbar and a power of 10 W. These samples were generally thermally stable so that no annealing was needed.\par

To vary the oxidation state of the metallic layer M, we either varied the thickness of this layer (in the case of natural oxidation) or the oxidation time (in the case of plasma oxidation), keeping the thickness of layer M constant at 1.6 nm in the last case. In the case of natural oxidation, a capping layer of Pt was deposited on top of these trilayers in order to prevent them from further oxidation. In the case of plasma oxidation, the capping layer was not necessary because the oxidation state was particularly stable due to the thick M layer (1.6 nm).\par

\section{Magnetic Properties\label{s:4}}

\subsection{Natural oxidation}

\begin{figure}
	\centering
		\includegraphics[width=10cm]{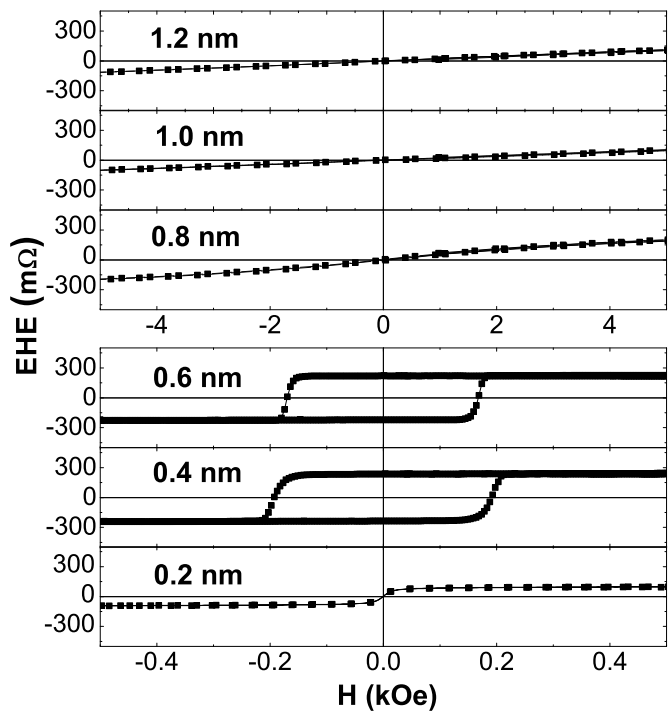}
		\caption{Hall resistance as a function of the applied field for Pt/Co/Al($d$)+Ox/Pt samples, for various Al thickness $d$. Oxidation is carried out in a 160 mbar O$_2$ atmosphere for 30 min. The field is applied perpendicular to the plane of the trilayers.}
	\label{fig:naturelle}
\end{figure}

The magnetic characterization of these samples was performed by extraordinary Hall effect (EHE) in a standard four-probes Hall geometry\cite{canedy, monso}. This technique gives direct information on the perpendicular component $M_z$ of the sample magnetization. Figure \ref{fig:naturelle} shows the hysteresis loops obtained at room temperature for a Pt(3 nm)/Co(0.6 nm)/Al($d$)+Ox/Pt(2 nm) stack with natural oxidation for various Al thicknesses $d$.

For thick Al layers ($d\geq$0.8 nm), the magnetization loops show no hysteresis and zero remanence. Magnetization saturates at high external field (|H|>1 kOe), where the perpendicular component of the Co magnetization $M_z$ no longer depends on H and the residual slope comes from the ordinary Hall contribution \cite{monso}. This demonstrates that the out-of-plane direction is a hard axis.\par

For $d$=0.6 nm and $d$=0.4 nm, the magnetization loop exhibits a square hysteresis with sharp magnetization reversal and a coercive field of 200 Oe. These features indicate that the out-of-plane direction becomes an easy axis. Finally, at $d$=0.2 nm, the magnetization loop shows no hysteresis, zero remanence and its amplitude has decreased, which can be attributed to partial oxidation of the Co layer.\par

These results demonstrate that, due to oxidation of the Al layer, the magnetic anisotropy of the Co layer is strongly modified. To correctly understand the EHE measurements, one should remind that three sources contribute to the magnetic energy of the thin Co layer: (i) the interfacial anisotropies (Pt/Co and Co/AlOx), (ii) the shape anisotropy, (iii) the stress induced anisotropy. The magnetization state of the Co layer arises from the competition between these three quantities.\par

The initial slope of the loop measured by EHE gives information about the resulting out-of-plane anisotropy. For thick and thin Al layers ($d\geq$0.8 nm and $d\leq$0.2 nm), the slope at zero field is small and the out-of-plane direction is a hard axis. For intermediate thicknesses (0.8 nm >$d$>0.2 nm), the EHE signal is a square loop, the interfacial Co/AlOx and Pt/Co PMA dominate the other anisotropy sources and the initial slope is nearly infinite. At his stage, we cannot determine accurately the magnetization state of the samples with thick ($\leq$0.8 nm) or thin ($\geq$0.2 nm) Al layers, since the EHE signals due to in-plane magnetization or to out-of-plane magnetization forming domain structures have the same shape (no remanence).\par

\begin{figure}
	\centering
		\includegraphics[width=15cm]{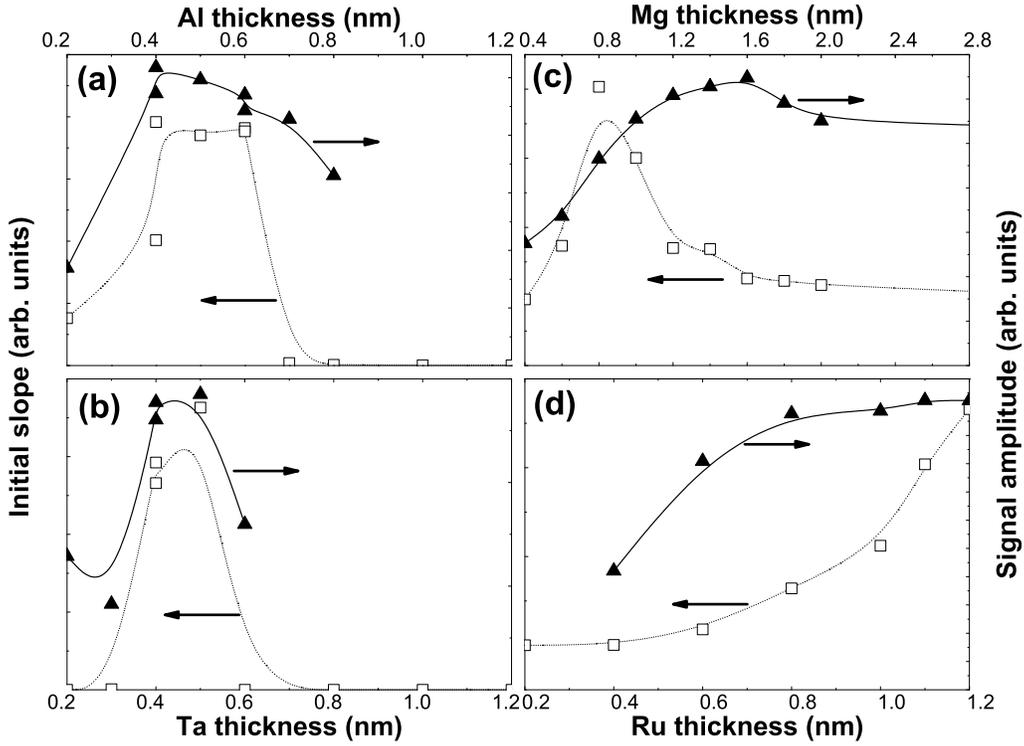}
		\caption{Initial slope and amplitude of the hysteresis loops measured by EHE for Pt/Co/MOx/Pt samples under natural oxidation, with M=Al, Ta, Mg, Ru. The dotted and solid lines are guides for the eyes.}
	\label{fig:al_nat}
\end{figure}

These results are summarized in Fig. \ref{fig:al_nat}. The initial slope and the magnitude of the EHE hysteresis loops are displayed as a function of the oxide thickness $d$, for various metallic layers M: Al, Mg, Ta, Ru. In all these samples, the initial slope (open squares) can be decomposed in three zones: for thick layers, the slope is small, for intermediate thicknesses (the range of which depends on the metal) the slope increases until a maximum, and finally decreases again for thin layers. For M=Al, Ta, Mg, the magnetization loop has a square hysteresis when the slope reaches a maximum, which indicates that the magnetization is out-of-plane. Concerning M=Ru, the slope decreases with decreasing thickness. This means that the maximum of perpendicular magnetic anisotropy appears for Ru layers thicker than 1.2 nm.\par

Another characteristic reported on Fig. \ref{fig:al_nat} is the amplitude of the EHE resistance, associated with the magnitude of the $M_z$ component. This amplitude shows a plateau for thick layers and decreases when the layers become thinner. The onset of decrease of the signal amplitude generally corresponds to the maximum of the initial slope, except for M=Mg where the magnitude begins to decrease before the anisotropy reaches its maximum. This decrease in EHE signal indicates a reduction of the thickness of the magnetic layer: the oxygen atoms penetrate in the bulk Co after the anisotropy has reached a maximum. The case of Mg is more difficult to analyse since in this case, the decrease in EHE amplitude is less abrupt, probably due to an inhomogeneous oxidation.\par

\subsection{Plasma oxidation}

\begin{figure}[width=8cm]
	\centering
		\includegraphics{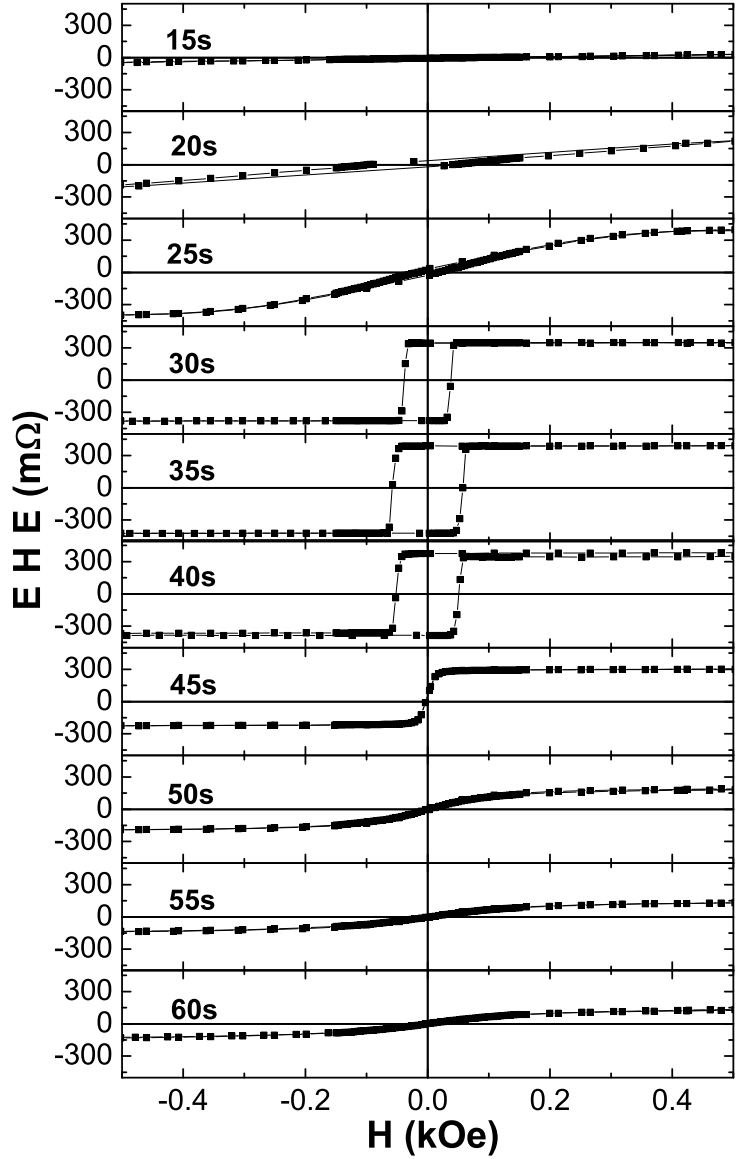}
		\caption{Hall resistance as a function of the applied field for Pt/Co/Al($t$)+Ox/Pt samples under plasma oxidation, for various oxidation times. The field is applied perpendicular to the plane of the trilayers.}
	\label{fig:plasma}
\end{figure}

\begin{figure}
	\centering
		\includegraphics[width=10cm]{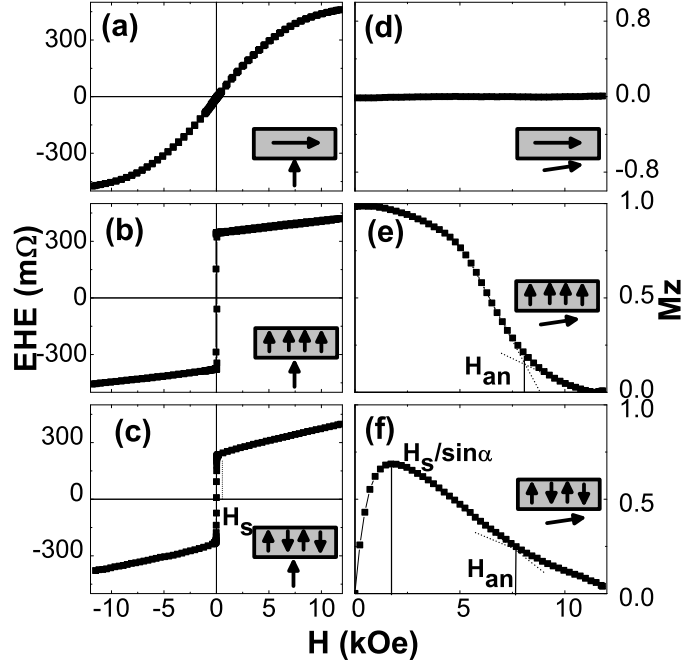}
				\caption{Extraordinary Hall effect with perpendicular field for three characteristic oxidation times: $t$=15s (a), $t$=30s (b) and $t$=45s (c); Extraordinary Hall effect with planar field for three characteristic oxidation times: $t$=15s (d), $t$=30s (e) and $t$=45s (f).}\label{fig:EHE_planaire}
\end{figure}

To understand the origin of this anisotropy enhancement, we prepared Pt(3 nm)/Co(0.6 nm)/Al(1.6 nm)+Ox($t$) samples. These samples were deposited as described in section \ref{s:3} and oxidized in a $O_2$ plasma, for various oxidation times $t$. Fig. \ref{fig:plasma} presents the EHE signal as a function of the oxidation time.
For samples with $t\leq$25s, the magnetization loops show no hysteresis and nearly zero remanence. Magnetization saturates at high field (|H|>1 kOe, see Fig.\ref{fig:EHE_planaire}(a)). This demonstrates that the out-of-plane direction is a hard axis. Note however that the magnetization loops for samples with $t=$20s and $t=$25s are open which clearly indicates that the magnetization is out-of-plane forming a domain structure.\par
\begin{figure}
	\centering
		\includegraphics[width=8cm]{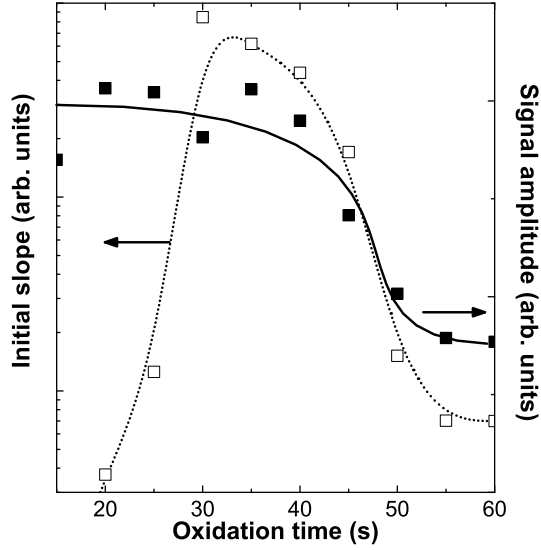}
		\caption{Initial slope - logarithmic scale - and amplitude of the hysteresis loops measured by EHE for Pt/Co/AlOx samples under plasma oxidation as a function of the oxidation time. The dotted and solid lines are guides for the eyes.}
	\label{fig:Al_plasma}
\end{figure}

For 30s$\leq t\leq$40s, the samples show a square hysteresis loop with sharp magnetization reversal and a non-zero coercivity. Finally, from $t\geq$45s, the EHE measurements show no hysteresis and zero remanence. The magnitude of EHE decreases for $t\geq$45s, which indicates that the effective thickness of the magnetic layer decreases. This is interpreted as due to the oxidation of the Co layer, as previously discussed. Initial slopes and signal amplitudes are displayed in Fig. \ref{fig:Al_plasma}. They show the same trends as in Fig. \ref{fig:al_nat}: the magnetization loops for a decreasing thickness of Al (natural oxidation) possess the same characteristics than the magnetization loop for an increasing oxidation time (plasma oxidation).\par

To determine the magnetic state of the Co layer, we performed EHE measurements applying the external magnetic field in the plane of the layers. Fig. \ref{fig:EHE_planaire} shows the EHE signal for small ($t$=15s), intermediate ($t$=30s) and long oxidation time ($t$=45s), for perpendicular (left figures) and in-plane (right figures) external field. In this last configuration, the sample plane is misaligned from the direction of the external field by a small angle $\alpha$ (of the order of 1 or 2°), in order to induce a coherent rotation of the magnetization. For these measurements, the samples have been previously saturated in positive perpendicular field (when the remanent magnetization is $M_z=1$) or demagnetized (when the remanent magnetization is less than 1 and when the magnetization loop is open). The Hall resistance is then measured in increasing field from 0 to 12 kOe, the maximum available field, then normalized to the amplitude of the magnetization loop measured in the perpendicular configuration, in order to obtain the magnitude of $M_z$.\par
For $t$=15s, the variation of $M_z$ as a function of the planar applied field is very small (see Fig.\ref{fig:EHE_planaire}(d)), indicating that at zero field, the magnetization lies in the plane of the sample. This is confirmed by SQUID measurements (not shown).\par

For $t$=30s, the magnetization is perpendicular with a 100\% remanence at zero field. In this case, the $M_z$ component decreases from 1 to 0 under the external planar field and saturates for applied field of 8 to 9 kOe (see Fig.\ref{fig:EHE_planaire}(e)).\par

\begin{figure}
	\centering
		\includegraphics[width=8cm]{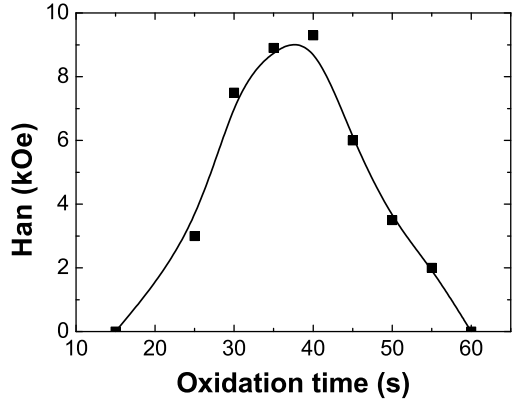}
				\caption{Out-of-plane anisotropy field of Pt/Co/AlOx samples, measured by EHE, as a function of the oxidation time.}\label{fig:anis}
\end{figure}

Finally, at $t$=45s, the $M_z$ component, initially zero, reaches a maximum at $H\approx 2$ kOe (see Fig.\ref{fig:EHE_planaire}(f)), corresponding to a projection $H_z=H\sin\alpha\approx50$ Oe, which coincides with the saturation field $H_s$, measured in Fig.\ref{fig:EHE_planaire}(c) in perpendicular applied field. This demonstrates that the magnetization structure at zero field corresponds to an out-of-plane magnetic domain structure. The same behaviour was observed for the samples oxidized during $t=$15s and $t$=60s. The magnetic behaviour strongly depends on the oxidation degree and can be summarised as follows:
\begin{itemize}
\item $t\leq$15s: the magnetization of the Co layer lies in the plane
\item 20s$\leq t\leq$25s: the magnetization of the Co layer lies out-of-plane, forming a domain structure
\item 30s$\leq t\leq$40s: the magnetization of the Co layer lies out-of-plane, forming a single domain structure
\item 45s$\leq t$: the Al layer is over-oxidized, oxygen penetrates in the Co layer (the thickness of the magnetic layer decreases). The balance between the anisotropy field and the demagnetizing field induces the magnetization of the Co layer lying out-of-plane in a domain structure
\end{itemize}
Fig. \ref{fig:anis} presents the out-of-plane anisotropy field measured by EHE with in-plane external field. Setting $M_x=\sqrt{1-M_z^2}$, one can determine the field at which the $M_x$ magnetization saturates at $M_x=1$, corresponding to the anisotropy field $H_{an}$. Similarly to the initial slope in Fig. \ref{fig:Al_plasma}, the anisotropy field reaches a maximum of 8-9 kOe for an oxidation time of the order of 30-40s.\par
In a recent publication, Lacour {\it et al.} \cite{lacour} studied the effect of oxidation on similar samples grown on a Ta buffer layer. The authors obtained out-of-plane Co magnetization even without oxidation and they only observed the second part of the anisotropy crossover: from the out-of-plane monodomain magnetic state to a superparamagnetic state. The superparamagnetic state they obtain for long oxidation times is consistent with our out-of-plane domain structure. However, they claimed strong differences between their results and ours. The fact that they obtain 100\% remanence (out-of-plane magnetization) even for zero oxidation time shows that their samples exhibit a larger PMA than ours, probably because of the Ta buffer layer. Furthermore, the authors focused on the coercive field of their samples and the lack of anisotropy measurements doesn't allow one to draw any definite conclusions about the variation of the magnetic anisotropy with oxidation time.\par

\section{Chemical analysis using X-ray spectroscopy\label{s:5}}

The above EHE measurements (Fig. \ref{fig:plasma} and Fig. \ref{fig:EHE_planaire}) show a clear dependence of the magnetization behaviour on oxidation, but do not allow us to comment on the oxidation state of the Al and Co layers. For $t\geq$45s, the amplitude of the EHE signal decreases and this can be attributed to the penetration of O atoms in the Co layer. However, nothing can be said about the smaller oxidation times.\par

To understand the role of oxygen in the modification of the magnetic anisotropy, we performed X-ray Absorption Spectroscopy (XAS) and X-ray Photoelectron Spectroscopy (XPS) experiments at the Advanced Photoelectric-effect Experiments (APE) beamline of the ELETTRA synchroton in Trieste (Italy). The aim of these element-specific measurements is to correlate the changes in macroscopic magnetic properties with changes in average chemical composition at the Co/Al interface. For the Co thickness used here (0.6 nm), XAS gives information on the chemical composition averaged over the Co layer, while XPS provides information on the electronic structure of Co at the Co/AlOx interface mainly (the collected intensity is exponentially decreasing with a penetration depth of the order of 0.5-0.6 nm, so that we expect that the main contribution comes from the top Co-Al bondings and Co-Co bondings between the first and second Co monolayers).\par

\subsection{X-ray Absorption Spectroscopy}

\begin{figure}
	\centering
		\includegraphics[width=16cm]{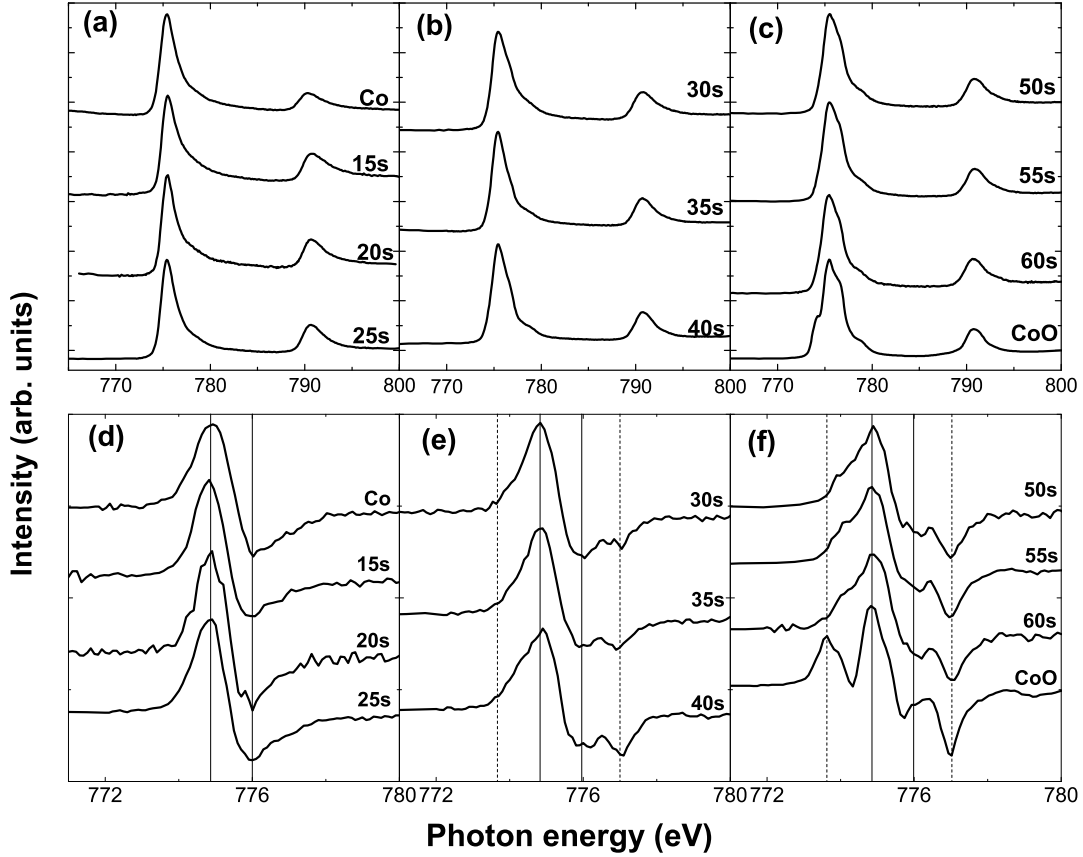}
		\caption{Co $L_{2,3}$ XAS spectra of the Pt/Co/AlOx trilayers for under-oxidized (a), optimally oxidized (b) and over-oxidized (c) samples and their corresponding derivatives (d,e,f) at Co $L_2$ edge. Spectra of Co and CoO are given for reference in (a,d) and (c,f), respectively. The solid and dotted vertical lines indicate the energy position of the shoulders in the XAS spectra due to Co and CoO, respectively.}
	\label{fig:XAS}
\end{figure}

XAS measurements were performed scanning the X-ray energy around the Co $L_{2,3}$ absorption edges (760-800 eV). Absorption spectra were obtained for perpendicular incidence, by measuring the drain current via the sample holder. Figure \ref{fig:XAS} shows the XAS spectra of the Pt/Co/AlOx trilayers for different oxidation times. Pure Co and CoO XAS spectra are given for reference in Fig. \ref{fig:XAS}(a) and (c), respectively.\par
For $t\leq$25s (Fig. \ref{fig:XAS}(a)), the spectra resemble the ones of pure Co. For 30s$\leq t\leq$40s (Fig. \ref{fig:XAS}(b)), shoulders appear in the absorption spectra indicating the presence of small contributions due to CoO. For 45s$\leq t\leq$60s (Fig. \ref{fig:XAS}(c)), these shoulders become very clear and the absorption spectra strongly resemble the ones of pure CoO.\par

Fig. \ref{fig:XAS}(c,d,f) shows the corresponding derivatives of the XAS spectra which reveal more clearly the presence of CoO in the Co layer. The vertical lines indicate the energy of the peaks of the derivative, due to Co (solid line) and CoO (dashed line). For small oxidation time (Fig. \ref{fig:XAS}(d)), only peaks corresponding to pure Co are present. For intermediate oxidation times (Fig. \ref{fig:XAS}(e)), a small peak appears at 777 eV, which corresponds to a small contribution of CoO. Finally, for long oxidation times (Fig. \ref{fig:XAS}(f)), two CoO peaks are present, but still smaller than the pure Co peaks. Note the energy shift of the derivative peaks as a function of the oxidation time, attributed to the modification of the chemical environment. This clearly indicates that the Co layer is free from oxygen atoms for small oxidation times, then a small oxidation occurs, probably near the interface, and finally, oxygen deeply penetrates inside the Co layer.\par

\subsection{X-ray Photoelectron Spectroscopy\label{s:6}}

\begin{figure}[t]
	\centering
		\includegraphics[width=8cm]{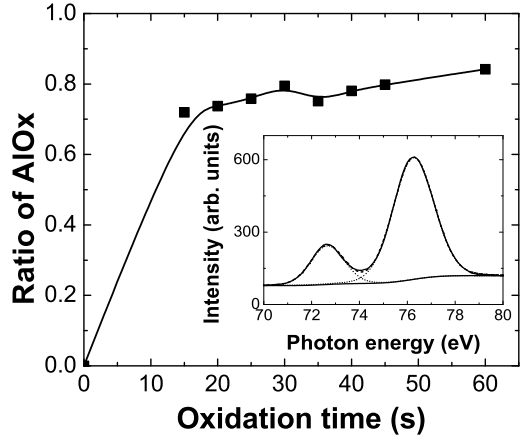}
		\caption{Ratio of oxidized aluminium estimated from Al $3p$ X-ray photoelectron spectra. The solid line is a guide for the eyes. Inset: XPS spectrum of Al 3$p$ with its Shirley background (solid line) for a sample oxidized during $t$=30s. The dotted lines are the gaussian fits of the XPS peaks and the dashed line is the sum of these estimated peaks.}\label{fig:Al3p}
\end{figure}

In order to have a more quantitative estimate of the chemical composition of Al and Co, we performed X-ray photoelectron spectroscopy on the 3$p$ levels of Al and the 2$p$ levels of Co. The beamline energy is set at $h\nu=765eV$ to probe the Al 3$p$ levels. The inset of Fig. \ref{fig:Al3p} shows a typical spectrum of Al 3$p$ in a sample oxidized during 30s. We observe two peaks located at 72.67 eV (corresponding to pure Al) and at 76.27 eV (corresponding to pure AlOx). Actually, we expect two peaks located at 72.9 eV and 72.5 eV, corresponding to the Al 3$p_{1/2}$ and 3$p_{3/2}$ levels respectively. However, the energy resolution of the beamline and the width of these peaks do not allow separating the two peaks.\par
The ratio of AlOx present in the layer is determined as $R=A_{AlOx}/(A_{AlOx}+A_{Al})$, where $A_{AlOx(Al)}$ is the estimated area of the AlOx (Al) peaks (the areas are estimated by gaussian fits - see inset of Fig. \ref{fig:Al3p}). Fig. \ref{fig:Al3p} shows that aluminium is not fully oxidized, even after 60s of plasma oxidation. However this measurement is restricted to the first half of the Al layer (the aluminium thickness is 1.6 nm, whereas the estimated escape length is smaller than 1 nm) and indicates that nearly 70\% of the first nanometer of the aluminium layer is oxidized during the first 15s of plasma oxidation. For longer oxidation times, the oxygen penetrates deeply inside the Al layer towards the Co/AlOx interface.\par
\begin{figure}[t]
	\centering
		\includegraphics[width=8cm]{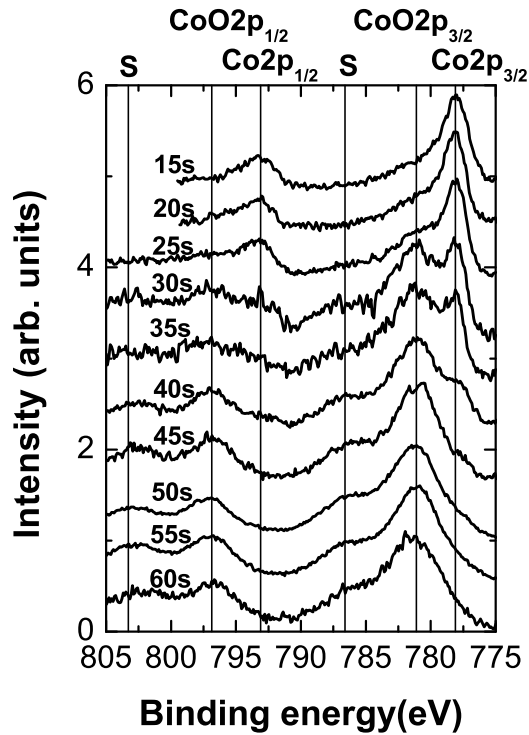}
		\caption{XPS spectra of Co 2$p$ for Pt/Co/Al($t$)+Ox samples as a function of the oxidation time $t$.}\label{fig:xps}
\end{figure}

To obtain information on the chemical composition of the Co layer, XPS is performed setting the beamline energy at $h\nu=1130eV$ to probe the Co 2$p$ levels. Fig. \ref{fig:xps} shows Co 2$p$ spectra obtained for the ten samples. The spectrum of the $t$=60s sample corresponds to pure CoO \cite{CoO}: one can distinguish two main peaks, corresponding to CoO 2$p_{1/2}$ and CoO 2$p_{3/2}$ core levels (lying at 796.3 eV and 781.1 eV respectively) and two satellite peaks (denoted S and lying at 803.3 eV and 786.7 eV resp.) which arise from the charge transfer between O 2$p$ and Co 3$d$. This charge transfer allows the coexistence of two types of transitions: $2p^63d^7\rightarrow 2p^53d^7+e^-$ (satellite peaks) and $2p^63d^8L\rightarrow 2p^53d^8L+e^-$ (main peaks), where $L$ is a hole brought by oxygen \cite{CoO}. The main peaks are shifted towards larger binding energies with respect to Co metal peaks, due to oxygen environment (similarly to AlOx peak). The spectrum of the $t$=15s sample is similar to that of pure cobalt \cite{sicot}, with Co 2$p_{1/2}$ and Co 2$p_{3/2}$ lying at the binding energies of 792.5 eV and 778.1 eV respectively.\par
\begin{figure}[t]
	\centering
		\includegraphics[width=8cm]{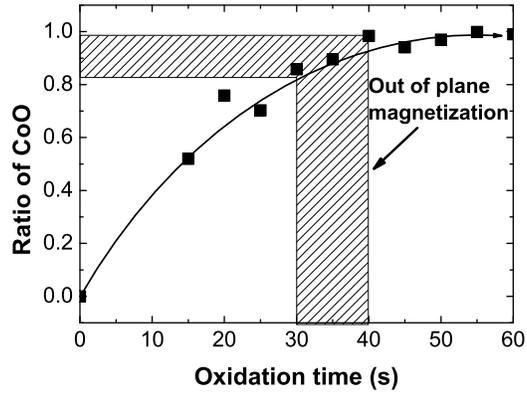}
		\caption{Ratio of oxidized cobalt estimated from Co 2$p$ X-ray photoelectron spectra, as a function of the oxidation time.}
	\label{fig:Co2p}
\end{figure}

It clearly shows that Co is unoxidized for $t$=15s. At $t$=20s  and $t$=25s, a small contribution of Co-O bondings appears. Between $t$=30s and $t$=40s, both Co-O and Co-Co bondings coexist and for oxidation times longer than 45s, the Co contribution completely desappears. The ratio of CoO in the Co layer has been estimated in Fig. \ref{fig:Co2p}, from gaussian fits of the XPS spectra (not shown).\par
The spectra clearly show that the Co/AlOx interface composition changes from quasi-pure Co to pure CoO during the oxidation process. Since XPS measurements are surface sensitive, the spectra give information about the Co chemical composition averaged over the Co layer, with a dominant contribution from the topmost Co monolayer (top Co-Al bondings, bottom and lateral Co-Co bondings). For samples oxidized during $t\leq$15s (in-plane anisotropy), the spectra have mainly a pure Co character. For $t$=20s and $t$=25s, the relatively high amount of estimated CoO (40\% to 60\%), compared to XAS results (undetectable CoO), suggests an inhomogeneous presence of oxygen atoms limited to the interface, which only allow small enhancement of the magnetic anisotropy (see Fig. \ref{fig:anis}).\par

For $t$=30s to $t$=40s, the spectra present both Co and CoO characters, showing that most of the Co located at the Co/AlOx interface possesses Co-O bondings (Co-Al bondings are replaced by Co-O bondings), while the small contribution of pure Co in the XPS spectra arises from the remaining deep Co-Co bondings. The estimated ratios (>80\%) show that the interface is nearly fully oxidized. Finally, from $t$=45s to $t$=60s, the Co character disappears which indicates that the first Co monolayer (at least) is totally oxidized (Co-Al and Co-Co bondings are replaced by Co-O bondings) and the CoO ratio reaches 100\%.\par

\section{Conclusion\label{s:7}}

\begin{figure}[t]
	\centering
		\includegraphics[width=8cm]{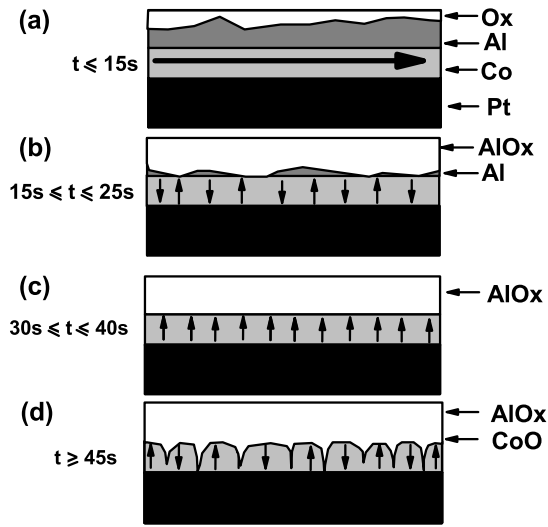}
	\caption{Schematics of the influence of the oxidation on the magnetic properties of the Co layer in a Pt/Co/AlOx trilayer: (a) for very small oxidation times, the magnetization lies in the plane; (b) for small oxidation times, the magnetization lies out-of-plane forming domain structure; (c) for optimal oxidation times, the magnetization lies out of plane in a single domain structure; (d) for long oxidation times, the magnetization lies out-of-plane forming domain structure.}\label{fig:interp}
\end{figure}

The above results strongly suggest that the onset of PMA is related to the appearance of a significant density of interfacial Co-O bondings at the Co/AlOx interface. These results are summarized in Fig. \ref{fig:interp}. For small oxidation times (Fig. \ref{fig:interp}(a)), the Al layer is only oxidized at 70\% and the magnetization of the Co layer lies in the plane. At $20s\leq t\leq25s$ (Fig. \ref{fig:interp}(b)), the perpendicular magnetic anisotropy increases due to the progressive appearance of oxygen at the Co/AlOx interface and the magnetization lies out-of-plane forming a domain structure.\par

For intermediate (or optimal) oxidation times (Fig. \ref{fig:interp}(c)), 80\% of the Al layer is oxidized and an important amount of oxygen atoms have reached the Co/Al interface, so that most of the Co-Al bondings are replaced by Co-O bondings. This indicates a non homogeneous oxidation: oxygen atoms diffuse through the grain boundaries of the Al layer\cite{bae} (AlOx is amorphous but Al is polycrystalline), so that they reach the Co/AlOx interface before Al is fully oxidized. In this case, the magnetic anisotropy at Co/AlOx interface reaches a maximum: the Co/Pt and Co/AlOx PMA overcome the other anisotropy sources and the magnetization of the Co layer lies out-of-plane in a single domain structure.

Finally, for long oxidation times (Fig. \ref{fig:interp}(d)), the Co layer is being oxidized through the grain boundaries \cite{bae}. The penetration of oxygen atoms in this way decreases the exchange coupling between the magnetic grains in the Co layer, reducing the energy of the magnetic domains. Consequently, the Co magnetization breaks into perpendicular magnetic domains, pointing upward or downward, probably forming labyrinthic or stripe domain structures \cite{kaplan}.

In conclusion, we have shown that interfacial oxidation of Co in Pt/Co/AlOx trilayers is at the origin of strong perpendicular magnetic anisotropy. Perpendicular anisotropy crossover induced by annealing in such structures has also been studied \cite{apl}. Ab-initio calculations, aiming at clarifying the role of Co-O hybridization in magnetization anisotropy energy, are presently underway.\par

\begin{acknowledgments}
The authors acknowledge very fruitful discussions with C. Lacroix, A. Bergmann and M. Chschiev. We are also grateful for the experimental support and assistance of B. Pang, L. Ranno, O. Fruchart and P. Torelli.
\end{acknowledgments}

\end{document}